\documentclass[12pt]{article}
\usepackage{a4wide,epsf,epsfig}
\newcommand{\xji}{\bar{x}}

\begin{document}
\begin{flushright}
WU B 00-21 \\
hep-ph/0101059
\end{flushright}

\begin{center}
\vskip 3.5\baselineskip
{\Large\bf The skewed quark distribution of the pion \\[1ex]
      at large momentum transfer} 
\vskip 2.5\baselineskip
C. Vogt
\vskip \baselineskip
Fachbereich Physik, Universit\"at Wuppertal, 42097 Wuppertal,
         Germany 
\vskip \baselineskip

\vskip 2.5\baselineskip

\textbf{Abstract} \\[0.5\baselineskip]

\parbox{0.9\textwidth}{
We derive an explicit and model-independent representation for the skewed quark
distribution of the pion in the limit of large momentum transfer. Our result is
expressed in terms of the conventional pion distribution amplitudes and 
complies with known properties of skewed parton distributions such as symmetry
and polynomiality conditions. The continuity of the result at the 
points~$\pm\xi$ is manifest, whereas its derivatives exhibit discontinuities.}

\vskip 1.5\baselineskip
\end{center}

% end of title page

%{\bf Introduction.}
%\vskip\baselineskip
Skewed parton distributions (SPDs) have been introduced as parametrizations of
soft matrix elements of quark and gluon field operators between hadronic states
with different momenta~\cite{Mueller1994,Ji1997,Radyushkin1996}. These matrix 
elements are relevant to hard electroproduction processes such as, for 
instance, deeply virtual Compton scattering (DVCS). An interesting feature of 
SPDs is that they are interpolating functions between ordinary parton 
distribution functions and hadronic form factors. More precisely, in the 
forward limit SPDs are given by parton distribution functions and form factors
are moments of SPDs.

Because of the nonperturbative nature of the corresponding soft hadronic matrix
elements information about the form of SPDs has to be taken from experiment 
or, alternatively, one may resort to various 
models~\cite{JiSong1997}-\nocite{Petrov,Anikin2000,DFJK1,
Radyushkin1998,Afanasev1998,Bakulev2000}\cite{CV2000}. 
In Refs.~\cite{DFJK3} and~\cite{BDH2000} an exact overlap representation of 
SPDs in terms of light-cone wave functions has been given.

In the present work we derive an explicit and model-independent expression for
the skewed quark distribution of the pion at large momentum transfer $t$. 
To this end, we consider
DVCS off pions. The amplitude of this process is known to factorize into a hard
photon-parton scattering and a SPD~\cite{Mueller1994,Ji1997,Radyushkin1996}.
If we additionally demand that $s,-t,-u\gg\Lambda^2$, where $\Lambda$ is a
typical hadronic scale of the order of 1 GeV, the process is amenable to a 
perturbative analysis within the
hard scattering approach of Ref.~\cite{LB1979}. The amplitude then factorizes
into two pion distribution amplitudes (DAs) and a hard scattering amplitude
which, in lowest-order quantum chromodynamics (QCD), is given in terms of 
one-gluon exchange diagrams.
In this way, the SPD is represented in terms of two pion DAs. We will show that
our result satisfies general properties of SPDs, is continuous, and its 
derivatives exhibit discontinuities at the points $\pm\xi$. 

The outline of the paper is the following. We start by briefly introducing 
our notation. We then present the derivation of the SPD of the pion at 
large~$t$ within the hard scattering approach. Since the derivation is 
completely analogous to that of the two-pion distribution amplitude from the 
process $\gamma^*\gamma\to\pi\pi$ at $\Lambda^2\ll s\ll Q^2$~\cite{DFKV1999}, 
which is related to DVCS by crossing, we will only repeat the essential steps 
and refer the reader to Ref.~\cite{DFKV1999} for the details. We reproduce the
general properties and elaborate on the behavior of our result at the points
$\pm\xi$. We conclude with our summary.

%{\bf Definitions.} 
\begin{figure}[h]
\begin{center}
\psfig{file=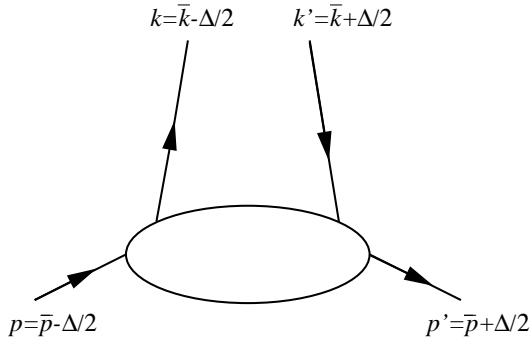,bb=20 260 580 610,width=8cm,angle=0}
\caption{Schematical picture of a SPD, represented by the blob.} 
                            \label{sketch}
\end{center}
\end{figure}
We denote the momenta of the incoming and outgoing pions by $p$ and $p'$, 
respectively (see, also, Fig.~\ref{sketch}).  In this work, we use Ji's 
parametrization~\cite{Ji1998} throughout, i.e., we introduce the average 
momentum $\bar{p}=(p+p')/2$, 
where from now on we follow the notation of Ref.~\cite{DFJK3} and employ the
``bar'' convention in order to define average momenta and average momentum 
fractions. Denoting the momentum transfer by $\Delta=p'-p$, we can write the 
skewedness parameter which describes the change in plus momentum as 
$\xi=-\Delta^+/(2\bar{p}^+)$. We define the average parton momentum 
$\bar{k}=(k+k')/2$, where $k\, (k')$ is the momentum of the emitted (absorbed) 
parton, and we introduce the average momentum fraction 
$\xji=\bar{k}^+/\bar{p}^+$.
With these conventions, we write the formal definition of a skewed quark
distribution for a flavor $q$ in terms of a nondiagonal hadronic matrix
element as~\cite{PolyakovWeiss1999} 
\begin{eqnarray}
 \int\frac{dz^-}{2\pi}\,e^{i\xji\bar{p}^+z^-}\,\langle \pi(p')|
 \bar{\psi}_q(-z^-/2)\gamma^+\psi_q(z^-/2)|\pi(p)\rangle 
  = 2\, H_q(\xji,\xi,t),  \label{def spd}
\end{eqnarray} 
where $z^-$ in the quark fields is a short hand for $z=[0,z^-,{\bf 0}_\perp]$
in light-cone coordinates.

We now turn to a discussion of the Compton process 
$\gamma^*\pi^+\to\gamma\pi^+$ in the deeply virtual region, i.e., for 
$-t\ll Q^2$.
In analogy to the nucleon case~\cite{Ji1997,Radyushkin1996}, we can write the
helicity amplitude in leading order $\alpha_s$ and leading twist in terms of 
a hard part and the SPD $H_q(\xji,\xi,t)$:
\begin{eqnarray}
 {\cal M}_{\lambda\lambda'}&=&-\frac{1}{2}\,
   \delta_{\lambda\lambda'}\sum_q e_0^2 \, e_q^2 \int_{-1}^1 d\xji\,
   \bigg[\frac{1}{\xji-\xi+i\,\epsilon}+\frac{1}{\xji+\xi-i\,
   \epsilon}\bigg]\, H_q(\xji,\xi,t),        \label{helicity amplitude}
\end{eqnarray}
with $e_0$ being the positron charge and $\lambda$ and $\lambda'$ denote the
helicities of the virtual and real photon, respectively. The helicity selection
rule stems from the collinear scattering of massless quarks. Note that the 
above expression is independent of the photon virtuality~$Q^2$.

%{\bf The skewed quark distribution at large $t$.}
Demanding that $s,-t,-u \gg\Lambda^2$ while keeping the condition $-t\ll Q^2$
we can subject the process to a perturbative investigation within the hard 
scattering approach~\cite{LB1979}. The helicity amplitude may then be expressed
by a convolution of a hard scattering amplitude $T_H$ and two pion DAs 
$\phi_\pi$~\cite{LB1979}:
\begin{eqnarray}
 {\cal M}_{\lambda\lambda'}(s,t,Q^2)&=&\frac{f_\pi^2}{24}\int_0^1 dx \, dy \,
 \phi^*_\pi(y;\mu_F) \, T_H(x,y,s,t,Q^2) \, \phi_\pi(x;\mu_F),
\label{hsp amplitude}
\end{eqnarray}
where $f_\pi\simeq 131$ MeV is the pion decay constant and $x,\,y$ denote the 
momentum fractions of the quarks relative to their respective parent hadrons.
$\mu_F$ is the factorization scale.
In leading order $\alpha_s/t$ only the lowest Fock state contributes to 
Eq.~(\ref{hsp amplitude}).

\begin{figure}[ht]
\begin{center}
\psfig{file=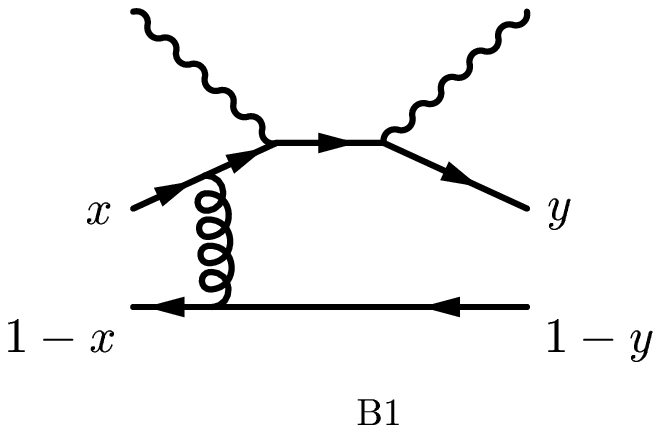,bb=200 640 400 790,width=6cm,angle=0}
\psfig{file=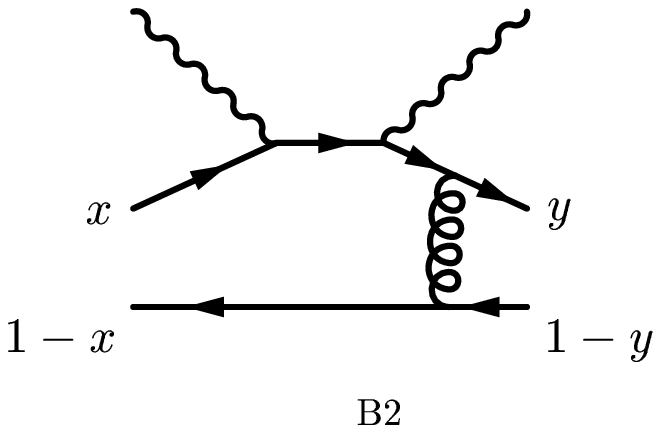,bb=180 640 380 790,width=6cm,angle=0}
\caption{Representatives of the of the handbag diagrams for 
         $\gamma^*\pi\to\gamma\pi$ in leading order using light-cone gauge.
         The remaining handbag diagrams are obtained by permutation of the
         photon vertices and of the fermion lines.}
\label{handbag diagrams}
\end{center}
\end{figure}
While the soft physics is embodied by the pion DAs, the hard scattering 
amplitude describes the hard photon-parton scattering. The relevant Feynman
diagrams and the corresponding expressions for $T_H$ are related by crossing
to those of the pion pair production process $\gamma^*\gamma\to\pi\pi$.
The diagrams of that process are explicitly given in Ref.~\cite{DFKV1999}. 
As we have discussed there it is useful to employ the gluon propagator in 
light-cone gauge. In this case, only particular diagrams contribute to the hard
scattering amplitude in leading order, namely, those handbag diagrams where the
gluon does not couple to the quark which connects the two photon vertices,
see Fig.~\ref{handbag diagrams}. 
The contributions of the other handbag diagrams where the gluon couples 
to the quark connecting the photon vertices and the contributions of the 
cat's-ears diagrams where the 
photons couple to different quark lines are suppressed by powers of $1/Q$.

The physical picture is then such that either before or after the hard 
photon-quark scattering the quarks exchange a gluon. This gluon is soft 
relative to the dominant hard scale $Q^2$ and is absorbed into the SPD, 
i.e., the gluon is assigned to the blob in Fig.~\ref{sketch}.

The expression for the hard scattering amplitudes can readily be obtained from 
those given in Ref.~\cite{DFKV1999}. One simply has to replace 
$s\leftrightarrow t$ and, since one of the outgoing pions in the pion pair
production becomes an incoming pion in the Compton process, the substitution 
$y\leftrightarrow 1-y$ has to be made\footnote{Note that the ``bar'' notation
of Ref.~\cite{DFKV1999} has a different meaning than in the present work:
in Ref.~\cite{DFKV1999} we use $\bar{y}=1-y$, etc.}, which exchanges the role
of quark and antiquark. We have explicitly checked
this crossing rule by recalculating the hard scattering amplitudes for DVCS.
In order to derive the representation of $H_q(\xji,\xi,t)$ we have to calculate
Eq.~(\ref{hsp amplitude}) and compare with Eq.~(\ref{helicity amplitude}). 
To this end, we rewrite $s$ and $u$ in terms of $\xi$ and $Q^2$. Up to 
corrections of order $t/Q^2$ we find:
\begin{equation}
 s=\frac{1-\xi}{2\xi}\,Q^2, \quad u=-\frac{1+\xi}{2\xi}\,Q^2. \label{su}
\end{equation}
Further, we express the integration variables $x$ and $y$ in terms of the 
average momentum fraction $\xji$, defined above, and $\xi$.
In the case where the photon couples to a $u$ quark, i.e., $q=u$, we have
\begin{eqnarray}
 y&=&\frac{\xji-\xi}{1-\xi}, \quad {\rm for\quad B1}, \label{y to xi}\\
 x&=&\frac{\xji+\xi}{1+\xi}, \quad {\rm for\quad B2}. \label{x to xi}
\end{eqnarray}
Since $0\leq x,y\leq 1$ it follows that 
\begin{eqnarray}
 \xi \leq \xji  \leq 1, &&\quad {\rm for\quad B1}, \label{restrict-b1}\\
 -\xi \leq \xji \leq 1, &&\quad {\rm for\quad B2}.
\end{eqnarray}
It is interesting to note that only diagram B2, where the gluon is exchanged
after the hard photon-quark scattering, contributes to the central region 
$-\xi\leq \xji\leq\xi$. This can be understood as follows. 

The returning quark of diagram B2, i.e., the quark between the photon and 
gluon vertices, is off-shell and has negative momentum fraction $\bar{x}-\xi$ 
in the region $-\xi\leq \xji\leq\xi$. Therefore, the quark may be interpreted 
as an antiquark with positive momentum fraction $-(\bar{x}-\xi)$ being emitted 
by the initial pion. But this just corresponds to the usual interpretation of 
a SPD in the central region: there, the SPD describes the emission of a 
$q\bar{q}$ pair. On the other hand, diagram B1 does not allow for such an 
interpretation because the gluon is exchanged before the photon-quark 
scattering, i.e., the emitted quark is off-shell, whereas the returning quark 
is on-shell and thus cannot have negative momentum fraction. Interpreting the 
emitted quark as an absorbed antiquark, one finds that the SPD describes the 
absorption of a $q\bar{q}$ pair. However, this situation is kinematically 
forbidden in DVCS and automatically excluded as a result of 
condition~(\ref{restrict-b1}).

After the above substitutions and convoluting the leading order hard scattering
amplitude with the pion DAs we obtain the helicity 
amplitude~(\ref{hsp amplitude}) in the limit $\Lambda^2\ll -t\ll Q^2$. For
$q=u$ we have
%\newpage
\begin{eqnarray}
 &&\hspace{-50pt} {\cal M}^u_{\lambda\lambda'}(\xi,t)=\frac{1}{2}\,
   \delta_{\lambda\lambda'}\,e_0^2e_u^2\,\frac{8 \pi f_\pi^2}{9}\nonumber\\
 &&\times \Bigg\{\int_{-\xi}^1 d\xji\, \bigg[
   \frac{1}{\xji-\xi+i\,\epsilon}+\frac{1}{\xji+\xi-i\,\epsilon}
   \bigg]\,\frac{1+\xi}{1-\xji} \, \phi_\pi\bigg(\frac{\xji+\xi}{1+\xi}
     \bigg) \,\tilde{I}(\xji,-\xi,t;\phi^*_\pi) \nonumber\\
 &&\hspace{10pt}+\int_\xi^1 d\xji \,\bigg[\frac{1}{\xji-\xi
   +i\,\epsilon}+\frac{1}{\xji+\xi-i\,\epsilon}\bigg]\,
   \frac{1-\xi}{1-\xji} \, \phi^*_\pi\bigg(\frac{\xji-\xi}{1-\xi}\bigg)\,
    \tilde{I}(\xji,\xi,t;\phi_\pi)
   \Bigg\},  \label{result hsp amplitude}
\end{eqnarray}
where the first term results from diagram B2 and the second term from B1.
The integral $\tilde{I}(\xji,\xi,t;\phi_\pi)$ is defined by 
\begin{equation}
 \tilde{I}(\xji,\xi,t;\phi_\pi)=\int_0^1 dy\, \frac{\alpha_s(\mu_R)}{t}\,
  \frac{(1-y)(1+\xi)+(\xji+\xi)}{y(1+\xi)-(\xji+\xi)}\,
  \frac{\phi_\pi(y)}{1-y}. \label{def integral}
\end{equation}
A natural choice of the renormalization scale $\mu_R^2$ is the gluon 
virtuality, which is typically of the order $t$. In the case of the photon 
coupling to the $\bar{d}$ quark, one has to make the replacements 
$x\rightarrow 1-x,\,y\rightarrow 1-y$ and $e_u^2\rightarrow e_d^2$ in the hard
scattering 
amplitudes for DVCS. The calculation shows that using the isospin symmetry of 
the pion DA, $\phi_\pi(x)=\phi_\pi(1-x)$, the helicity amplitudes for
$q=u$ and $q=\bar{d}$ are the same up to the charge factor. We find analogous
counting rules for the contributions of the various helicities of the photons
as in the case of the pion pair production, which is reflected by the 
Kronecker delta in Eq.~(\ref{result hsp amplitude}). Longitudinally polarized
virtual photons contribute with $1/(Q\sqrt{-t})$ and photons with opposite 
helicities contribute with $1/Q^2$.

By direct comparison of Eq.~(\ref{result hsp amplitude}) with 
Eq.~(\ref{helicity amplitude}) we can read off the expression for the 
skewed $u$-quark distribution of the pion:
\begin{eqnarray}
 H_u(\xji,\xi,t)=-\frac{4}{9}\pi f_\pi^2\Bigg\{\hspace{-20pt}&&
  \Theta(\xji+\xi)
  \frac{1+\xi}{1-\xji} \, \phi_\pi\bigg(\frac{\xji+\xi}{1+\xi}\bigg) \,
  \tilde{I}(\xji,-\xi,t;\phi^*_\pi)  \nonumber\\
  +&&\Theta(\xji-\xi)\frac{1-\xi}{1-\xji} \, 
  \phi^*_\pi\bigg(\frac{\xji-\xi}{1+\xi}\bigg) \,
  \tilde{I}(\xji,\xi,t;\phi_\pi)\Bigg\}.          \label{pert pion spd}
\end{eqnarray}
According to our remark below Eq.~(\ref{def integral}) concerning the case
$q=\bar{d}$ it is clear that $H_{\bar d}(\xji,\xi,t)=H_u(\xji,\xi,t)$.
With the relation between skewed quark and antiquark distributions 
$H_{\bar q}(\xji,\xi,t)=-H_q(-\xji,\xi,t)$ (see, e.g., Ref.~\cite{DFJK3}), 
we find
\begin{equation}
 H_d(\xji,\xi,t)=-H_u(-\xji,\xi,t). \label{d spd}
\end{equation}
for the skewed $d$-quark distribution. From the suppression of higher Fock
states it immediately follows that 
\begin{equation}
 H_s(\xji,\xi,t)\equiv 0.
\end{equation}

The symmetry under $\xi\leftrightarrow -\xi$ \cite{Ji1998} is manifest in our 
result~(\ref{pert pion spd}). 
In order to show that expression~(\ref{pert pion spd}) complies with the 
polynomiality condition~\cite{JiSong1997,Radyushkin1999} we proceed 
analogously as in the
case of the two-pion distribution amplitude~\cite{DFKV1999}. We replace the 
integration variable $y$ appearing in the integral 
$\tilde{I}(\xji,-\xi,t;\phi_\pi)$ in the first term of 
Eq.~(\ref{pert pion spd})
by $x$ and revert the substitutions~(\ref{y to xi}) and (\ref{x to xi}).
Using the binomial expansion and the relation 
$(a^n-b^n)/(a-b)=\sum_{i=1}^n a^{n-i}\, b^{i-1}$ this gives
\begin{eqnarray} 
 &&\hspace{-20pt}\int_{-1}^1 d\xji \,\xji^{n-1}\, H_u(\xji,\xi,t) =
 -\frac{4}{9}\pi f_\pi^2 \int_0^1 dx\, dy \, \frac{\alpha_s}{t}
 \frac{\phi_\pi(x)}{1-x}\frac{\phi^*_\pi(y)}{1-y} \nonumber \\
 &&\times
 \Bigg[-2\sum_{\scriptstyle i=0 \atop \scriptstyle {\rm even}}^{n-1}\bigg( 
   \begin{array}{c} n-1 \\ i \end{array} \bigg)\,x^{n-1-i}\, [ (1-x)\xi ]^i
  +2\xi\sum_{\scriptstyle i=1 \atop \scriptstyle {\rm odd}}^{n-1}\bigg( 
   \begin{array}{c} n-1 \\ i \end{array} \bigg)\,x^{n-1-i}\, [ (1-x)\xi ]^i
  \nonumber\\ 
 &&\hspace{135pt}-(1-\xi^2)\sum_{i=1}^{n-1} [y+(1-y)\xi]^{n-1-i}\,
        [x-(1-x)\xi]^{i-1} \Bigg].                  \label{spd moments}
\end{eqnarray}
Obviously, the right hand side is a polynomial of order $n$ in $\xi$. Using 
relation~(\ref{d spd}) we find for the first moment, i.e., $n=1$, the 
pion form factor in the collinear hard scattering approximation~\cite{LB1979}:
\begin{eqnarray} 
 \int_{-1}^1 d\xji\, H_u(\xji,\xi,t) =
 \frac{8 \pi f_\pi^2}{9} \int_0^1 dx\, dy \, \frac{\alpha_s}{t}
 \frac{\phi_\pi(x)}{1-x}\frac{\phi^*_\pi(y)}{1-y}.
\end{eqnarray}

\begin{figure}[t]
\begin{center}
\psfig{file=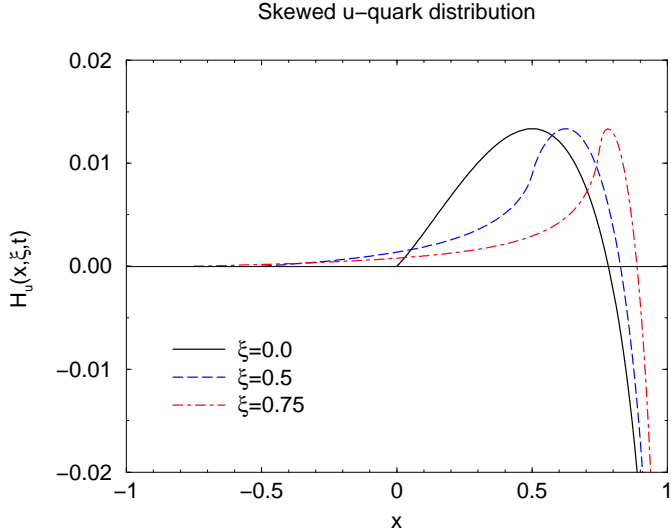,bb=45 70 560 675,width=7cm,angle=-90}
\caption{The SPD (\ref{as pion spd}) at different values of $\xi$  
         and $-t=10$~GeV$^2$.}   \label{spd plot}
\end{center}
\end{figure}

Phenomenological investigations of the $\pi$-$\gamma$-transition form factor
strongly indicate that the pion DA is close to its asymptotic form
$\phi_{\rm as}(x)=6 x (1-x)$ \cite{Musatov}. Taking this as an example
the skewed $u$-quark distribution reads
\begin{eqnarray}
 H_u(\xji,\xi,t)=-16\pi f_\pi^2\frac{\alpha_s}{t}\Bigg\{&&\hspace{-18pt}
  \Theta(\xji+\xi)\,\frac{\xji+\xi}{1+\xi}\bigg[1+\frac{\xji
  -\xi}{1-\xi}\ln\bigg|\frac{1-\xji}{\xji-\xi}\bigg|\bigg] \nonumber\\
  +&&\hspace{-18pt}\Theta(\xji-\xi)\,\frac{\xji-\xi}{1-\xi}\bigg[
  1+\frac{\xji+\xi}{1+\xi}\ln\bigg|\frac{1-\xji}{\xji+\xi}\bigg| 
  \bigg]\Bigg\}. \label{as pion spd}
\end{eqnarray}
The result is shown in Fig.~\ref{spd plot}. As we can see the skewed quark
distribution is logarithmically divergent at the point $\xji=1$. This 
divergence has the same origin as the divergence of the perturbative limit of 
the two-pion DA at 
$z=\zeta$~\cite{DFKV1999} and reflects the breakdown of the collinear hard 
scattering approach at the endpoints where the virtualities of the internal 
partons, in particular the one of the gluon, become small. For a detailed
qualitative and quantitative discussion of this issue see Ref.~\cite{DFKV1999}.

Only little is known about the behavior of SPDs at the points $\xji=\pm\xi$. 
It has been pointed out in Ref.~\cite{Radyushkin1996} that the continuity of 
SPDs at these points is necessary for the factorization in DVCS and other hard
electroproduction processes. Inspection of expression~(\ref{as pion spd})
shows that our result is indeed continuous at $\xji=\pm\xi$. Its derivatives at
$\xji=\pm\xi$, however, do not exist. This means that our result for the skewed
quark distribution at large $t$ is an example of a SPD, the derivatives of 
which manifestly have singularities at these points.
It can easily be seen that this is 
not only true in the case where we utilize the asymptotic form of the pion DA 
in Eq.~(\ref{pert pion spd}). The pion DA has an expansion in terms of 
Gegenbauer polynomials~\cite{LB1979} and using an arbitrary finite number of 
terms in this expansion one always finds singularities of the derivatives at 
$\xji=\pm\xi$. Note that SPDs which result from integrating double 
distributions show the same behavior at these points~\cite{Radyushkin1999}.
For a discussion of the complete twist-2 structure of double distributions 
see Ref.~\cite{PolyakovWeiss1999}.

Finally, we remark that the determination of the factorization scale of the 
pion DAs as well as that of the SPD requires an analysis of QCD corrections to
the process amplitude. Such an analysis is, however, beyond the scope of this 
work.

\vskip\baselineskip

To summarize, we have used the hard scattering approximation to perturbatively
ana\-lyze DVCS off pions in the limit $\Lambda^2\ll-t\ll Q^2$, where $\Lambda$
is of ${\cal O}(1$ GeV). Within this approach, we have derived an explicit and
model-independent representation of the skewed quark distribution of the pion 
at large momentum transfer $t$ in terms of pion DAs. Employing
light-cone gauge we have found that the dominant contributions to the process
come from the usual handbag diagrams, whereas contributions from gluons 
attached to the hard part
of the diagram as well as contributions from the cat's-ears diagrams are 
suppressed by powers of $1/Q$. We have shown that our result is fully 
consistent with the symmetry relation and the polynomiality condition and 
verified that the first moment of the SPD reproduces the well known expression
for the pion form factor in the collinear hard scattering approximation. 
Furthermore, the SPD is continuous and its derivatives are discontinuous at 
the points $\pm\xi$.

\vskip\baselineskip

{\bf Acknowledgements.} 
I would like to thank M. Diehl, D. M\"uller, R. Jakob, P.~Kroll, and B.~Postler
for useful discussions. Financial support by the Deutsche 
Forschungs\-gemeinschaft is acknowledged.

\newpage

\vskip3\baselineskip

\begin{center}
{\bf Erratum: Skewed quark distribution of the pion at large momentum transfer
\\[1ex] Phys.\ Rev.\ D {\bf 64}, 057501 (2001) [hep-ph/0101059]}

\vskip2\baselineskip

C.Vogt
\end{center}

\vskip2\baselineskip

In expressions~(10) and (14)--(16), the Mandelstam variable $t$ should 
be replaced with $(-t)$. 
In Eq.~(11), the argument of the pion distribution amplitude $\phi_\pi^*$ 
in the second term in curly brackets should read $(\bar{x}-\xi)/(1-\xi)$
instead of $(\bar{x}-\xi)/(1+\xi)$.  

A further typo is in Eq.~(16), where the first term in the square
brackets is to be replaced with 1/2. The correct expression thus reads
\begin{eqnarray}
 H_u(\xji,\xi,t)= -16\pi f_\pi^2\frac{\alpha_s}{(-t)} \Bigg\{&&\hspace{-18pt}
  \Theta(\xji+\xi)\,\frac{\xji+\xi}{1+\xi}\bigg[\frac12+\frac{\xji
  -\xi}{1-\xi}\ln\bigg|\frac{1-\xji}{\xji-\xi}\bigg|\bigg] \nonumber \\
  +&&\hspace{-18pt} \Theta(\xji-\xi)\,\frac{\xji-\xi}{1-\xi}\bigg[
  \frac12+\frac{\xji+\xi}{1+\xi}\ln\bigg|\frac{1-\xji}{\xji+\xi}\bigg| 
  \bigg]\Bigg\} \ . \nonumber
\end{eqnarray} 
Consequently, the plot of the skewed quark distribution as shown in 
Fig.~3 is incorrect. 

The above result for $H_u(\xji,\xi,t) $ agrees with the 
result reported in Ref.~\cite{ji:2003}, where a different convention 
for the pion decay constant has been used with $f_\pi \simeq 93$~MeV, 
whereas in the present work we use $f_\pi \simeq 131$~MeV. 

\vskip\baselineskip
{\bf Acknowledgments.}
I would like to thank M. Diehl and X. Ji for correspondence.

%\bibliography{apsamp}

\end{document}